\documentclass[onecolumn,showpacs,prd]{revtex4}
\usepackage{CJK}                     
\usepackage[dvips]{graphicx}
\usepackage{bm}                      
\usepackage{mathptmx}                
\usepackage{amssymb}  
\usepackage{dcolumn}
\usepackage{epsfig}
\usepackage{hyperref}
\usepackage{graphicx}
\usepackage{longtable}

\usepackage[dvips]{color}

\usepackage[normalem]{ulem}  

\newcommand{\NPA}[3]{Nucl.\ Phys.\ {\bf A #1}, #2 (#3)}
\newcommand{\NPB}[3]{Nucl.\ Phys.\ {\bf B #1}, #2 (#3)}

\newcommand{\PLB}[3]{Phys.\ Lett.\ B\ {\bf #1}, #2 (#3)}

\newcommand{\PRL}[3]{Phys.\ Rev.\ Lett.\ {\bf #1}, #2 (#3)}

\newcommand{\PRC}[3]{Phys.\ Rev.\ C\ {\bf #1}, #2 (#3)}
\newcommand{\PRD}[3]{Phys.\ Rev.\ D\ {\bf #1}, #2 (#3)}
\newcommand{\JPG}[3]{J.\ Phys.\ G\ {\bf #1}, #2 (#3)}

\begin{document}

\title{Color-flavor locked strange quark matter in a strong magnetic field}

\author{
Xin-Jian Wen \footnote{E-mail address: wenxj@sxu.edu.cn}  }

\affiliation{ Department of Physics and Institute of Theoretical
Physics, Shanxi University, Taiyuan 030006, China
              }
\date{\today}

\begin{abstract}
The quark quasiparticle model is extended to study the properties of
color-flavor locked strange quark matter at finite chemical
potential and in a strong magnetic field. We present a
self-consistent thermodynamic treatment by employing a chemical
potential dependent bag function. It is found that the magnetized
color-flavor-locked (MCFL) matter is more stable than other phases
within a proper magnitude of magnetic field. The stability window is
graphically shown for the MCFL matter compared with ordinate
magnetized matter. The anisotropic structure of MCFL matter is
dominated by the magnetic field and almost independent of the energy
gaps. A critical maximum magnetic field of about $1.56\times
10^{18}$ G is found, under which MCFL matter is absolutely stable
with respect to nuclear matter.

\end{abstract}

\pacs{24.85.+p, 12.38.Mh, 21.65.Qr, 25.75.-q}

\maketitle

\section{Introduction}
\label{sec:intro} Properties of strange quark matter (SQM) at low
temperature and high density regime have received a lot of
theoretical attentions. The stability of SQM in the interior of a
compact star has been investigated by many works
\cite{Witten84,Schertler1997,Buballa99}. The most symmetric pairing
state, color flavor locked (CFL) state, can only be realized if the
mass split between light quarks and strange quarks is small and/or
the chemical potential $\mu$ satisfies $\mu\gtrsim m_s^2/2\Delta$
\cite{alford04} with $m_s$ being the strange quark mass and $\Delta$
the pairing gap. CFL state is predicted to be more stable than
ordinary strange quark matter. So a hybrid star is suggested to be
constructed with color superconducting quark matter in the interior
and nuclear matter in the surface \cite{alford2001074017}. Recently,
the special properties of neutron star matter \cite{Dunc92} and
dense quark matter in the presence of a strong magnetic field have
aroused a lot of interest \cite{Chakra1996,Ebert03}. One may
naturally wonder how the external magnetic field affects the
stability condition and the equations of state of CFL matter.

In the past years, the magnetized strange quark matter (MSQM) and
CFL matter have been studied in the bag model \cite{Felipe05} and
improved in the Nambu-Jona Lasinio (NJL) model
\cite{Mizher10,Faya11,Avan11}. The CFL state has a wide range of
model parameters characterized by the so-called stability window
\cite{paul08}. In an external strong magnetic field, the magnetized
color flavor locked matter (MCFL) is expected to have interesting
properties \cite{Manuel07,Felipe11,Felipe2011b}, where the magnetic
field contribution is considered as the vacuum term in the
thermodynamical potential \cite{schwinger}. Due to the effect of the
strong magnetic field, the anisotropic pressure changes the
mechanical stability condition of strange quark matter, which is
especially important for the finite size spherical drop of SQM
\cite{Chakra1996,Felipe}. In literature, there are two kinds of
treatments with the chemical potential relation between various
components of the quark matter. One is the weak equilibrium for bulk
matter, such as the general treatment of $u$, $d$, and $s$ quarks in
quark stars. The other is non-$\beta$ equilibrium proposed in a
first order deconfinement transition of hadron matter to color
superconductivity. When the phase transition is unavoidable, the
non-$\beta$-stable intermediate state is assumed to have a short
life time \cite{Lugones05}.

In literature, the quark quasiparticle model has been developed by
many authors in studying the dense quark matter at finite
temperature and density \cite{schert98,Romat03,zheng2005}. The
importance of the quasiparticle model is the introduction of the
medium-dependent quark mass scale in describing QCD nonperturbative
properties. A widely accepted quark mass scale is derived at the
zero-momentum limit of the dispersion relations from an effective
quark propagator by resuming one-loop self-energy diagrams in the
hard dense loop approximation \cite{Schertler1997}. In the scale,
the quark mass is a function of its chemical potential and the
temperature. So the pressure and energy functions of SQM have
corrections by an additional bag function. As is known, the system
pressure and various component chemical potentials are very
important quantities for the equations of state of SQM. To balance
the electrostatic repulsion of electrically charged particles,
Poincare assumed a pressure to exist inside the particles
\cite{Einstein,TDLee}. The pressure is exerted by the physical space
outside the bag in the bag model. It has also been shown through the
dynamical quark mass in the NJL model \cite{NJL}. In the quark
quasiparticle model, it is presented by a chemical-potential-
dependent bag function, which is derived by the self-consistency
condition of thermodynamics. Recently, we have applied this model to
study the strangelets \cite{wen2009} and magnetized SQM with
interesting results \cite{wen2012}. The aim of this paper is to
investigate the special properties of the CFL state in an external
strong magnetic field. We derive a chemical potential and
field-strength- dependent bag function. Then we show the self
consistency of the thermodynamic treatment. Our results show that
the pairing effect together with an external magnetic field in a
proper magnitude can generally enlarge the stability window of MCFL
matter.

This paper is organized as follow. In Sec.~\ref{sec:thermo}, we
derive the self-consistent thermodynamic formulas of CFL and MCFL
matter in the quasiparticle model with and without the magnetic
field. Based on the running coupling constant between quarks, we
investigate the stability of different phases of SQM. Specially, the
stability windows of MSQM and MCFL matter are shown. In
Sec.~\ref{sec:num}, we show the numerical results and have a
discussion on the stability and the anisotropic structure of SQM in
strong magnetic field. The last section is a short summary.

\section{CFL matter and MCFL phase in the quasiparticle with a new
effective bag function}\label{sec:thermo}
\subsection{Thermodynamic treatment of CFL matter in the quasiparticle model}
We begin with the thermodynamical potential density of unpaired
quarks. At zero temperature, the thermodynamic potential density
$\Omega_\mathrm{free}$ is \cite{alford2001074017,Lugones02,Peng06}
\begin{eqnarray}\Omega_\mathrm{free}=\sum_{i=u,d,s} \frac{d_i}{48 \pi^2}
\Bigg\{p_f\bigg[3(2p_f^2+m_i^2)\sqrt{p_f^2+m_i^2}-8\mu_ip_f^2
\bigg]-3m_i^4 \ln\frac{p_f+\sqrt{p_f^2+m_i^2}}{m_i} \Bigg\},
\end{eqnarray}
where is the degeneracy factor $d_i$ is 6 for quarks and $d_i=2$ for
electrons. All the thermodynamic quantities can be derived from the
characteristic function by obeying the self-consistent thermodynamic
relation \cite{wen2009}.

In order to write down the expression for the thermodynamic
potential density of three-flavor quark matter in the color flavor
locked phase, as usually done, we assume that all nine (three
flavors times three colors) quarks have a common gap $\Delta$. To
order $\Delta^2$, the pairing contribution to the thermodynamic
potential density of color-flavor locked SQM can be written as
\cite{alford2001074017}
\begin{eqnarray}\Omega_\mathrm{CFL}=\Omega_\mathrm{free} +\Omega_\mathrm{pair},
\end{eqnarray}
where the pairing term from one quasiparticle with gap $\Delta_1$
and eight quasiparticles with gap $\Delta_2$ is \cite{Shov05}
\begin{eqnarray}
\Omega_\mathrm{pair}&\simeq&-\left (\frac{\bar{\mu}}{2 \pi}\right
)^2 (\Delta_1^2 +8\Delta_2^2)\label{eq:pair}\\ &\simeq& -3\Delta^2
\bar{\mu}^2/\pi^2.
\end{eqnarray}
where the approximate relation $\Delta_1=2\Delta_2=2\Delta$ is used
between the singlet and octet gaps. The mean chemical potential is
$\bar{\mu}=(\mu_u+\mu_d+ \mu_s)/3$. The common Fermi momentum $p_f$
is obtained by solving
\begin{eqnarray}\sum_i
\sqrt{p_f^2+{m_i^*}^2}=3\bar{\mu}.\label{Fermi}
\end{eqnarray}

In this paper, the effective quark mass in the framework of the
quasiparticle model is taken to be
\cite{Schertler1997,Schertler1997jpg,Pisarski1989},
\begin{eqnarray}m_i(\mu_i)=\frac{m_{i0}}{2}+ \sqrt{\frac{m^2_{i0}}{4}+
\frac{g^2\mu_i^2}{6\pi^2}}\, ,\label{mass1}
\end{eqnarray}
where $m_{i0}$, $\mu_i$, and $g$ are, respectively, the quark
current mass, quark chemical potential, and the strong interaction
coupling constant. In principle, the strong coupling constant is
running \cite{shir1997,wen2010}, and one can use a phenomenological
expression such as \cite{Patra1996},
\begin{equation}\label{g_run}
g^2(T=0,\mu_i)=\frac{48 \pi^2}{29}\left[\ln(\frac{0.8
\mu_i^2}{\Lambda^2})\right]^{-1}.
\end{equation}
However, to make the expressions simple and understandable, one can
adopt a constant $g$ with value in the range of (0, 5), as was done
in the previous literature \cite{Schertler1997}. Through the
following sections and figures, we choose the running coupling value
in the calculations. The current mass can be neglected for up and
down quarks. The strange quark current mass can be adopted as 120
MeV. If the vanishing current mass is assumed for up and down
quarks, Eq.~(\ref{mass1}) can be reduced to the simple form
\begin{eqnarray}m_i=\frac{g \mu_i}{\sqrt{6}\pi}.
\end{eqnarray}

The pressure  and the energy density for the CFL matter are,
respectively, given by
\begin{eqnarray}P&=&-\Omega_\mathrm{CFL}-B^*,\label{press}\\
E &=&\Omega_\mathrm{CFL}+\sum_i \mu_i n_i +B^*. \label{energy}
\end{eqnarray}
where the effective bag function $B^*$ is introduced as a residual
interaction energy density. To meet the thermodynamic consistency
condition, the $B^*$ function can be divided into two parts:
$\mu_i$-dependent part and the definite integral constant, i.e.,
$B^*=\sum_iB_i(\mu_i)+B_0$ (i= $u$, $d$, and $s$) where $B_0$ is
similar to the conventional bag constant. The concrete expression of
the chemical potential dependence of the effective bag function
$B^*$ is to be self-consistently derived in the next paragraph. The
advantage of this particular implementation is to show the automatic
confinement characteristic in the model.

The quark number density of the component $i$ is given as
\begin{equation}n_i=\frac{p_f^3}{\pi^2}+ \frac{2
\Delta^2\bar{\mu}}{\pi^2}.
\end{equation}
The total baryon number density $n_b$ is the sum of $n_i$ divided by
3, i.e., $n_b=(n_u+n_d+n_s)/3$. Eq.~(\ref{Fermi}) explicitly means
that all the three flavors of quarks have the same number density,
i.e. $n_u = n_d = n_s$. Therefore, the CFL phase is naturally
neutralized, and the number density of electrons is thus zero.
Accordingly, the weak equilibrium condition
$\mu_u+\mu_e=\mu_d=\mu_s$ becomes $\mu_u = \mu_d = \mu_s$ due to the
zero chemical potential of electrons.

 In order to satisfy the fundamental thermodynamic equation
\begin{eqnarray}E=-P+\sum_i\mu_i\frac{\partial P}{\partial \mu_i},
\end{eqnarray}the following requirement must be satisfied ~\cite{Goren1995,Schertler1997},
\begin{eqnarray}\left(\frac{\partial P}{\partial
m_i}\right)_{\mu_i}=0. \label{dp=0}
\end{eqnarray}
Considering Eq.(\ref{dp=0}), we have the vacuum energy density
function $B_i(\mu_i)$ through the following differential equation,
\begin{eqnarray}\frac{\mathrm{d} B_i(\mu_i)}{\mathrm{d} \mu_i}\frac{\mathrm{d} \mu_i}{\mathrm{d} m_i}=-\frac{\partial \Omega_\mathrm{CFL}}{\partial
m_i}.\label{eq:dBdm}
\end{eqnarray}
If the current mass of light quarks is neglected, one can integrate
Eq.~(\ref{eq:dBdm}) under the condition $B_i(\mu_i=0)=0$ and have
\begin{eqnarray} \frac{dB^*_{u, d}}{dm_i}=-\frac{d_i}{4\pi^2} \bigg[m_ip_f
\sqrt{m_i^2+p_f^2}- m_i^3
\ln(\frac{p_f+\sqrt{p_f^2+m_i^2}}{m_i})\bigg].
\end{eqnarray}

The effective bag constant including the effect of pairing was
investigated in the CFL matter by Ref.~\cite{Glugones04}. Here we
derive the effective bag function by self-consistent requirement
from Eq.~(\ref{eq:dBdm}). The pairing contributions is included in
the thermodynamic potential density. If we apply the running
coupling $g(\mu)$, the expression of the effective bag function is
changed into
\begin{eqnarray}\label{Bexp1}
B_i(\mu_i)&=&-\int^{\mu_i}_{\mu_i^c} \left.\frac{\partial
\Omega}{\partial
m^*_i}\right|_{T=0,\mu_i}\frac{\mathrm{d} m_i}{\mathrm{d}\mu_i} \mathrm{d}\mu_i \nonumber \\
&=&-\frac{d_i}{4\pi^2} \int^{\mu_i}_{\mu_i^c}\bigg[m_ip_f
\sqrt{m_i^2+p_f^2}+ m_i^3
\ln(\frac{p_f+\sqrt{p_f^2+m_i^2}}{m_i})\bigg] \frac{\mathrm{d}
m_i(\mu_i,g(\mu_i))}{\mathrm{d} \mu_i} d\mu_i,
\end{eqnarray}
where the lower limit $\mu_i^c$ of the integration over $\mu_i$
should satisfy $B_i(\mu_i^c)=0$. This requirement is equivalent to
including the integration constant into the $B_0$.

\begin{figure}[ht]
\centering
\includegraphics[width=7cm,height=7cm]{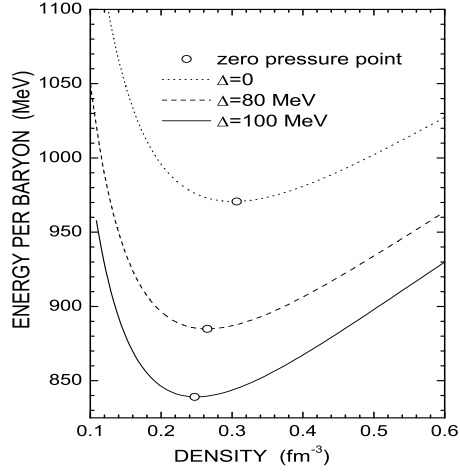}
\caption{The energy per baryon of ordinary quark matter and CFL
matter versus the baryon number density for $\Lambda=120$ MeV
without external magnetic field. The symbol ``$\circ$" denotes the
zero pressure equilibrium point of quark matter.
        }
\label{fig1}
\end{figure}
For the pairing energy gap, $\Delta=80$ MeV (dashed line) and
$\Delta=100$ MeV (solid line), the energy per baryon of CFL matter
versus the baryon number density is plotted in Fig. \ref{fig1}. On
the top of the panel, the $\Delta=0$ curve marked by a dotted line
represents the unpaired SQM. For larger pairing gap, the CFL matter
would have a smaller free energy per baryon. The minimum values of
the curves are located at the stable zero-pressure points marked by
open circles. This is the result of the thermodynamic
self-consistency requirement.

\subsection{MCFL phase in the present of the strong magnetic field}
To definitely describe the magnetic field of a compact star, we
assume a constant magnetic field ($H_{m,z}=H$) along the $z$ axis.
Due to the quantization of orbital motion of charged particles in
the presence of a strong magnetic field, known as Landau
diamagnetism, the single particle energy spectrum is \cite{Landau}
\begin{eqnarray}\varepsilon_i=\sqrt{p_z^2+m_i^2+e_iH(2n-\eta+1)},
\end{eqnarray}
where $p_z$ is the component of the particle momentum along the
direction of the magnetic field $H$, $e_i$ is the absolute value of
the electronic charge (e.g., $e_i=2/3$ for the up quark and 1/3 for
the down and strange quarks), $n=0,1,2,...$, is the principal
quantum number for the allowed Landau levels, and $\eta=\pm 1$
refers to quark spin-up and -down state, respectively. For the sake
of convenience, one usually sets $2\nu=2n-\eta+1$, where
$\nu=0,1,2,...$. The single particle energy then
becomes~\cite{Chakra1996}
\begin{eqnarray}\varepsilon_i=\sqrt{p_z^2+\bar{m}_{i, \nu}^2}.
\end{eqnarray}where $\bar{m}^2_{i,\nu}=m_i^2+2\nu e_iB$ denotes the square of
quark effective mass in the presence of a magnetic field.

In literature, the MCFL matter was successfully investigated in the
NJL framework \cite{Ferrer05,Ferrer06,Feng11,noron07,2008PRL}. Due
to the mixture of the photon field $A_\mu$ and the eighth component
$G_\mu^8$ of the gluon field, the original (u, d, s) quark
representation is divided into neutral, positively, and negatively
charged spinors with the quark rotated charges $\tilde{Q}$
\cite{alford00}. The $\tilde{Q}$ of different quarks are in units of
$\tilde{e}=e\cos{\theta}$ \cite{2008PRL}, where $\cos
\theta=g/\sqrt{e^2/3+g^2}$ \cite{0011333,shov04}. By considering the
relation between the magnitude of the applied external magnetic
field $H_\mathrm{ext}$ outside and the rotated field $\tilde{H}$
inside the color superconducting matter, one has
$\tilde{H}=H_\mathrm{ext}\cos \theta$. This can be simplified as
$\tilde{H}=H_\mathrm{ext}$ with $g\gg e$. In the following section,
we will use "magnetic field" in short when referring to the rotated
magnetic field $\tilde{H}$. The original symmetry pattern is
SU(3)$_{C+L+R}$ has been broken to the residual subgroup
SU(2)$_{C+L+R}$ \cite{noron07,Ferrer06}. The gap structure is
indirectly affected by the external field through the coupled gap
equations \cite{Ferrer06}. The original gap of CFL matter is split
into two different gaps, $\phi$ and $\Delta_\mathrm{H}$, in the
implementation of the MCFL matter. In principle, the gaps are
determined by the solution of the gap equation
\cite{Ferrer05,Ferrer06}. The gap $\phi$ is formed only by the pairs
of neutral quarks, while $\Delta_\mathrm{H}$ function has
contributions from quarks with opposite charges and neutral quarks
\cite{noron07}. By an analogy with the pairing contribution in
Eq.~(\ref{eq:pair}) for CFL matter, the total thermodynamic
potential density $\Omega_\mathrm{MCFL}$ is the sum of pairing term
and the thermodynamic potential $\Omega_\mathrm{MSQM}$ to lowest
nonzero order in $\Delta_\mathrm{H}/\mu_q$ at the quark chemical
potential $\mu_q=400\sim 500$ MeV,
\begin{eqnarray}\Omega_\mathrm{MCFL}=\Omega_\mathrm{MSQM}-\frac{\bar{\mu}^2}{\pi^2}(\phi^2+2\Delta_\mathrm{H}^2).
\end{eqnarray}
where $\Omega_\mathrm{MSQM}=\sum_i \Omega_i(m_i,\mu_i)$. This
approximation is obviously available to lowest nonzero order in
$\phi/\mu$ because $\phi$ could be smaller than $\Delta_\mathrm{H}$
\cite{Ferrer05,noron07}. The dependence of the gap
$\Delta_\mathrm{H}$ on the magnetic field can be understood through
the gap equation in NJL model \cite{Ferrer05,Ferrer06}. As is known
to us in the bag model, the quark interaction is simply included
into a single bag constant, which represents the energy density
difference. Therefore in order to obtain the gap value as the
function of the magnetic field, we can only resort to the
approximation solution from the gap equation in NJL model. For MCFL
matter, the gap parameters can be solved numerically through the gap
equation. In the strong-field limit, all the landau levels has been
drop down in addition to the lowest Landau level and the analytical
solution can be found in the equation,
\begin{eqnarray}1\approx \frac{g^2}{3\Lambda^2}\int_\Lambda
\frac{1}{\sqrt{(q-\mu)^2+2(\Delta_\mathrm{H})^2}}+
\frac{g^2\tilde{e}\tilde{H}}{6\Lambda^2} \int_{-\Lambda}^{\Lambda}
\frac{dq}{(2\pi)^2}
\frac{1}{\sqrt{(q-\mu)^2+(\Delta_\mathrm{H})^2}}.\label{eq:gap}
\end{eqnarray} For more detailed introduction, one can see Ref.
\cite{Ferrer05,Ferrer06}. Because the solution of Eq.~(\ref{eq:gap})
is associated with Landau energy level, the gap parameters show
oscillation behavior, known as the de Haas-van Alphen effect, as
long as $0.2<\tilde{e}\tilde{H}/\mu_q^2<0.5$. The oscillation ceases
when only the lowest Landau level contributes to the gap equations,
namely, $\tilde{e}\tilde{H}/\mu_q^2>1$ \cite{2008PRL}. In this case,
the analytical function was obtained for the dependence of the gap
on the magnetic field in Refs.~\cite{Ferrer05}. In the region
$\tilde{e}\tilde{H}/\mu_q^2 \sim 0.1$ we are interested in, the
parameters $\phi$ and $\Delta_\mathrm{H}$ show a narrow range of the
oscillation . The gap parameters with very weakly oscillation
behavior can be briefly supposed as constants in our investigation.
The thermodynamically self-consistence with this approach will be
shown in following section that zero pressure is obtained
simultaneously with the minimum of free energy. In the special case
of $\tilde{H}\sim 5\times 10^{17}$G, the gap parameters can be
approximately invariant. 

Because the ground state ($\nu=0$) is single degenerate while all
$\nu\neq 0$ states are doubly degenerate, we assign the spin
degeneracy factor ($2-\delta_{\nu 0}$) to the index $\nu$ Landau
level. We can do the calculations with the redefined charges of the
quarks in the 9-dimensional flavor-color representation
\cite{Ferrer05},
\begin{equation}\label{table-CFL}
\begin{tabular}{ccccccccc}
\hline \textrm{$s_b$}& \textrm{$s_g$}& \textrm{$s_r$}&
\textrm{$d_b$}& \textrm{$d_g$}& \textrm{$d_r$}& \textrm{$u_b$}&
\textrm{$u_g$}&
\textrm{$u_r$}\\
0 & 0 & $-1$ & 0 & 0 & $-1$ & $+1$ & $+1$ & 0\\
\hline
\end{tabular}
\end{equation}

 The thermodynamic potential density for charged quarks
at zero temperature is simplified to give
\begin{eqnarray}\Omega_i(m_i,\mu_i)&=&-\frac{d_i\tilde{e}_i\tilde{H}}{2\pi^2}\sum_{\nu=0}(2-\delta_{\nu
0})\int_0^{p_z} (\mu_i-\varepsilon_i)dp'_z \nonumber\\
&=&-\frac{d_i\tilde{e}_i
\tilde{H}}{4\pi^2}\sum_{\nu=0}^{\nu_\mathrm{max}}(2-\delta_{\nu
0})\Big\{ \mu_i \sqrt{\mu_i^2-\bar{m}^2_{i, \nu} }-\bar{m}^2_{i,
\nu}\ln(\frac{\mu_i + \sqrt{\mu_i^2-\bar{m}^2_{i,
\nu}}}{\bar{m}_{i,\nu}}) \Big\}.\label{omega0}
\end{eqnarray}where the $d_i=2$ only refers the spin degeneracy factor. The upper limit $\nu_\mathrm{max}$ of the summation
index $\nu$ can be understood from the positive value required by
the logarithm and square-root function in Eq. (\ref{omega0}). So we
have
\begin{equation}\nu\leq \nu_\mathrm{max}\equiv
\mathrm{int}[\frac{\mu_i^2-m_i^2}{2\tilde{e}_i\tilde{H}}],
\end{equation}where ``int" means the number before the decimal
point.

The quark number density is
\begin{eqnarray}n_i=\frac{d_i \tilde{e}_i\tilde{H}}{2\pi^2}\sum_{\nu=0}^{\nu_\mathrm{max}}(2-\delta_{\nu
0})\sqrt{\mu_i^2-\bar{m}_{i,\nu}^2}+ \frac{2
\bar{\mu}}{3\pi^2}(\phi^2+2\Delta_\mathrm{H}^2) .\label{eq:density}
\end{eqnarray}

In the preceding paragraphs, we only take into account the influence
of the magnetic field on the Landau energy levels of the particles.
In fact, the magnetic field pressure should be considered in
phenomenological presentations as in the previous work
\cite{Ferrer2011}, which should be distinguished with a
hydrodynamical pressure \cite{TDLee}. Due to the anisotropic
structure, the pressure is different in the directions parallel and
perpendicular to the field \cite{Strick2012}. By including the
matter contribution and the field contribution, the total parallel
pressure is given by
\begin{eqnarray}P_\parallel=-\Omega_\mathrm{MCFL}-\frac{\tilde{H}^2}{2}-B^*
\end{eqnarray}
and the transverse pressure is
\begin{eqnarray}P_\perp=-\Omega_\mathrm{MCFL}-M_f\tilde{H}+\frac{\tilde{H}^2}{2}-B^*,\label{eq:transpress}
\end{eqnarray}
where the system magnetization is $M_f=-(\partial
\Omega_\mathrm{MCFL}/\partial H)=\sum_i M_i$. The contribution from
the $i$th quark species reads
\begin{eqnarray}M_i=-\frac{\partial \Omega_i}{\partial \tilde{H}}= -\frac{\tilde{e}_i d_i}{2\pi^2}\sum^{\nu_\mathrm{max}}_{\nu=0} (2-\delta_{\nu
0})\int_0^{\sqrt{\mu_i^2-\bar{m}^2_{i,\nu}}}(\varepsilon_i-\mu_i+\frac{\nu
\tilde{e}_i\tilde{H}}{\varepsilon_i})dp_z. \label{eq:magnetization}
\end{eqnarray}
Substituting Eq.~(\ref{eq:magnetization}) into
Eq.~(\ref{eq:transpress}), one has the analytical expression,
\begin{eqnarray}P_\perp=\frac{d_i|\tilde{e}_i|^2
\tilde{H}^2}{2\pi^2}\sum_i\sum_{\nu=1}^{\nu_\mathrm{max}}\nu
\ln[\frac{\mu_i+\sqrt{\mu_i^2-\bar{m}^2_{i,\nu}}}{\bar{m}_{i,\nu}}]
+\frac{\tilde{H}^2}{2}-B^*.
\end{eqnarray}
The anisotropy of pressures in a strong magnetic field was recently
investigated by the bag model \cite{Felipe08} and the NJL model
\cite{Ferrer2010}. In our quasiparticle model, the quark effective
mass takes the place of the current mass in bag model to reflect the
medium effect in the high density environment. The pressure
splitting is
\begin{eqnarray}P_\perp-P_\parallel=-M_f\tilde{H}+\tilde{H}^2.
\end{eqnarray}
To guarantee the vanishing of both the parallel and transverse
pressure, one should require that $P_\parallel=0$ and $\partial
P_\parallel/\partial \tilde{H}=0$. It is obvious that the bag
constant should depend on the magnetic field
\cite{Felipe2011b,Ferrer2011}.

Similar to the approach ruled by the fundamental thermodynamic
relation in CFL matter, we have the following requirement,
\begin{eqnarray}\left(\frac{\partial P_\parallel}{\partial
m_i}\right)_{\mu_i,\tilde{H}}=0. \label{dpB=0}
\end{eqnarray}
Consequently, we have the following differential equation,
\begin{eqnarray}\frac{\mathrm{d} B_i(\mu_i)}{\mathrm{d} \mu_i}\frac{\mathrm{d} \mu_i}{\mathrm{d} m_i}=-\frac{\partial \Omega_\mathrm{MCFL}}{\partial
m_i}.\label{eq:dBdm}
\end{eqnarray}
We can express the effective bag function of MCFL matter as
\begin{eqnarray}B^*=\sum_i\sum_\nu B_{i,\nu}(\mu_i)+B_0,
\end{eqnarray} where the contribution of the component $i$ at the $\nu$ level is given
as
\begin{eqnarray}
B_{i,\nu}(\mu_i)&=&-\int^{\mu_i}_{\mu_i^c} \left.\frac{\partial
\Omega_\mathrm{MCFL}}{\partial
m^*_i}\right|_{T=0,\mu_i}\frac{\mathrm{d} m_i}{\mathrm{d}\mu_i} \mathrm{d}\mu_i \nonumber \\
&=&-\frac{d_i\tilde{e}_i\tilde{H}}{2\pi^2} \int^{\mu_i}_{\mu_i^c}
m_i
\ln(\frac{\mu_i+\sqrt{\mu_i^2-\bar{m}_{i,\nu}^2}}{\bar{m}_{i,\nu}})
\frac{\mathrm{d} m_i(\mu_i,g(\mu_i))}{\mathrm{d} \mu_i}
d\mu_i.\label{eq:B_nv}
\end{eqnarray}

For uncharged particles in the rotated representation, we can not
obtain discrete Landau levels and the integration of $\Omega_i$ can
be carried out to give \cite{wen2009}
\begin{eqnarray}\Omega_i 
&=&-\frac{d_i}{48\pi^2}
     \Bigg[
|\mu_i|\sqrt{\mu_i^2-{m_i}^2}\left(2\mu_i^2-5{m_i}^2\right)
  +3{m_i}^4\ln\frac{|\mu_i|+\sqrt{\mu_i^2-{m_i}^2}}{m_i}
    \Bigg].
\end{eqnarray} Correspondingly, the number density (\ref{eq:density})
and bag function (\ref{eq:B_nv}) for uncharged quarks should be
changed \cite{wen2009}.

\section{Numerical results: the stability and anisotropic structure of
MCFL}\label{sec:num}
\begin{figure}[ht]
\centering
\includegraphics[width=7cm,height=7cm]{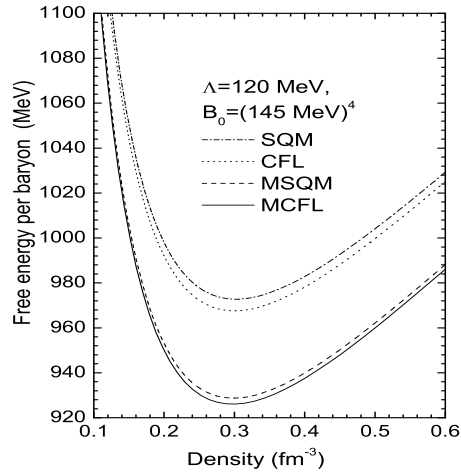}
\caption{The comparison of free energy per baryon versus the baryon
number density for MCFL,  MSOM, CFL, and SQM respectively. The
magnetic field strength and energy gaps are adopted as
$\tilde{H}=10^{17}$ G, $\phi=15$ MeV,  and $\Delta_\mathrm{H}=20$
MeV.
        }
\label{Fig2}
\end{figure}

It is known that different constitutions of three flavor quarks and
external environment will have different effects on the stability of
SQM. The stability for different phases was recently investigated by
an inequalities expression \cite{paul08}. In the framework of the
preceding quasiparticle model, we have done the numerical
calculations with the quark current mass values $m_u=5$ MeV,
$m_d=10$ MeV, and $m_s=120$ MeV. The constant term $B_0$ is $(145$
MeV$)^4$. In Fig. \ref{Fig2}, we give the energy per baryon for the
SQM, MSQM, CFL, and MCFL phases respectively. From the top to down
in the panel, there are obviously two features. Firstly, the quark
pairing effect greatly improves the stability of SQM. Secondly, the
external magnetic field in a proper magnitude lowers the free energy
per baryon through the rearrangement of Landau energy level of
magnetized quark matter. In a proper range of magnetic field
strength, we can get the general inequality relation of the energy
per baryon in the four kinds of quark matter phases as
\begin{eqnarray}\left.\frac{E}{A}\right|_\mathrm{MCFL}<\left.\frac{E}{A}\right|_\mathrm{MSQM}<\left.
\frac{E}{A}\right|_\mathrm{CFL}
<\left.\frac{E}{A}\right|_\mathrm{SQM}. \label{eq:rela}
\end{eqnarray}Here we should emphasize that the validity of the
comparison result in Eq.~(\ref{eq:rela}) will depend on the gap
value of CFL matter. If given a larger gap value $\sim
3\Delta_\mathrm{H}$, the CFL phase will be more stable than MSQM
phase.

\begin{figure}[ht]
\centering
\includegraphics[width=7cm,height=7cm]{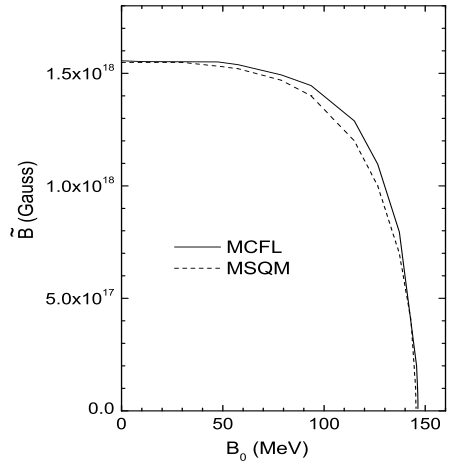}
\caption{Stability windows for MCFL matter and MSOM are denoted by
solid curve and dashed curve respectively. The parameters are the
same as Fig.~\ref{Fig2}. The two curves corresponds to the
borderline values 930 MeV of the energy per baryon.
        }
\label{Fig3}
\end{figure}

Now we investigate the range of the magnetic field strength required
by the stability condition of the MSQM and MCFL matter. As is known
that the effect of the magnetic field on the equation of state of
the magnetized SQM is remarkable \cite{wen2012}. Undoubtedly, the
influence of the magnetic field on the stabilities will depend on
its magnitude $\tilde{H}$. One may wonder how high is the field
magnitude is allowed in SQM according the stability hypothesis. The
possible strongest magnetic field in nature is generated either in
the interior of a rotating protoneutron star, or in the in
ultrarelativistic heavy ion collisions for a brief timescales of
order fm/c. The physical upper limit can be estimated as
$\tilde{H}_\mathrm{max}\sim 10^{20}$~G by assuming equivalent
uniform field and mass density under energy-conservation arguments
by applying the equipartition theorem on the magnetic energy and the
energy from quark matter \cite{Ferrer2010}. In this paragraph, we
will investigate the allowed field range by considering the
stability condition criterion. By comparing the energy per baryon
with that of the $^{56}$Fe nucleus (roughly 930 MeV), the stability
window ($E/n_b <930$ MeV) is obtained at zero pressure condition for
MSQM (dashed line) and MCFL (solid line) in Fig. \ref{Fig3}. For
MCFL matter, the two energy gaps are chosen as $\phi=15$ and
$\Delta_\mathrm{H}=20$ MeV. It is clearly seen that the pairing
energy can enlarge the stability window in the parameter space. The
maximum value of the magnetic field is allowed up to $H_m=1.56\times
10^{18}$ G in the absolute stability region, which is smaller than
that showed in Fig.5 of Ref.~\cite{Ferrer2011}. However, the special
values of the allowed maximum field are associated with the energy
gaps and the QCD scale parameter $\Lambda$.

\begin{figure}[ht]
\centering
\includegraphics[width=7cm,height=7cm]{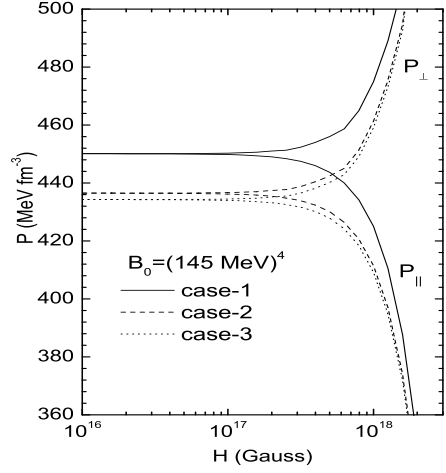}
\caption{The transverse pressure and parallel pressure of magnetized
CFL matter and unpaired matter as functions of the magnetic field
strength. The case-1, case-2, and case-3 of energy gaps ($\phi$,
$\Delta_\mathrm{H}$) are chosen as (30, 40), (15, 20), and (0, 0) in
unit of MeV, which are marked by solid, dashed, and dotted lines
respectively. The case-3 is the unpaired magnetized strange quark
matter.
        }
\label{Fig4}
\end{figure}

\begin{figure}[ht]
\centering
\includegraphics[width=7cm,height=7cm]{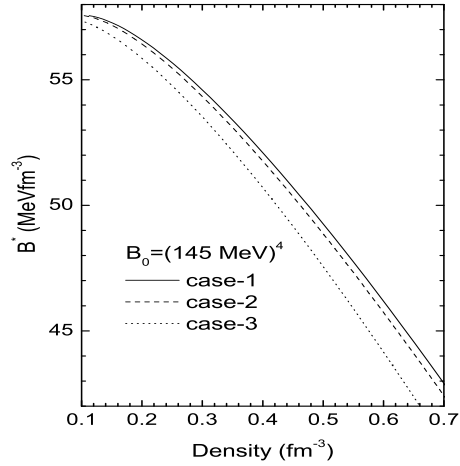}
\caption{The effective bag function $B^*$ versus the baryon number
density for different parameter sets of energy gaps. The field
strength is $\tilde{H}=10^{17}$ G. The other parameters are the same
as Fig.~\ref{Fig4}.
        }
\label{Fig5}
\end{figure}

The anisotropic structure was investigated recently in dense neutron
matter under the presence of a strong magnetic field \cite{Isay12}.
The difference of $P_\perp$ and $P_\parallel$ reflects the breaking
of the rotational symmetry by the magnetic field. The effect of
anisotropic pressure will become very important especially in
calculating the properties of strangelets, the spherical drop of
SQM. In contrast to the previous work about MCFL matter
\cite{Felipe11}, our equation of state in the quasiparticle model
implies $P_\parallel<P_\perp$ by including the field Maxwell
contribution. To satisfy the equilibrium condition of the self bound
matter, it is required that the parallel pressure and the transverse
pressure should vanish simultaneously $P_\parallel=P_\perp=0$
\cite{Ferrer2011}. A natural and realizable method is that the bag
parameter is field-dependent \cite{Felipe2011b}. Consequently, the
residual part of the bag constant in the present version of the
quasiparticle model should reflect its dependence on $\tilde{H}$
under the equilibrium condition. Just as pointed out in
Ref.~\cite{Ferrer2011}, one can consider a field-dependent bag
constant for gravitationally bound stars, because the internal
pressure produced by the magnetic field can be compensated by the
gravitational pressure. In Fig.~\ref{Fig4}, we show the parallel
$P_\parallel$ and the transverse $P_\perp$ pressures as functions of
the magnetic field strength. We call to the reader's attention that
our numerical result of Fig.~\ref{Fig4} is automatically obtained by
zero pressure condition instead of fixing a baryon number density
\cite{Isayev}. The solid and dashed lines represent the parameter
set ($\phi$, $\Delta_\mathrm{H}$) of the MCFL matter with (30, 40)
and (15, 20) in unit of MeV, respectively. The dotted line with
$\phi=\Delta_\mathrm{H}=0$ denotes the recovery of the unpaired MSQM
phase. The pressure lines feel a weak effect of the pairing gaps. It
is found that the pressure splitting is dominated by the magnetic
field and independent of the pairing energy gap. The magnetic field
strength has a common threshold value $\tilde{H}_\mathrm{th}\sim
3\times 10^{17}$~G for different gap values. When the magnetic field
strength is larger than the threshold value, the pressure anisotropy
starts to become noticeable: the transverse $P_\perp$ (or the
parallel $P_\parallel$) will increase (or decrease) rapidly far
beyond the constant values. Because the contribution of Maxwell term
to the pressure becomes dominant with the high value of the field
strength, the ascending branches as well as the descending branches
will become to overlap at high magnitude of the magnetic field. In
Fig.~\ref{Fig5}, we compare the effective bag function at the fixed
field strength $\tilde{H}=10^{17}$~G. It is obvious that the
residual interaction energy density $B^*$ will be decreasing with
the decreased distance between quarks, which is required by the
behavior of confinement potential energy. However at the same baryon
number density, the $B^*$ increases with the gap value of MCFL
matter. For ultrastrong field, i.e., $\tilde{e}\tilde{H}>\mu_q^2$,
the similar comparison can be out of the scope of the present work.

\section{Summary}
\label{Sec:conls} We have extended the quark quasiparticle model to
study the properties of both CFL matter and MCFL matter in an
external strong magnetic field. The self-consistent thermodynamic
treatment is obtained through an additional bag function, which is
dependent on the chemical potential and/or the magnetic field. It is
used to describe the residual interaction of quarks with the quark
effective masses in the quasiparticle model. The stability
properties of CFL and MCFL matter are calculated and compared with
ordinary SQM and MSQM. For a proper magnitude of the magnetic field,
about $\tilde{e}\tilde{H}<0.2 \mu_q^2$, the MCFL with smaller energy
per baryon is more stable than other phases of SQM. The magnetic
field makes both the paired and unpaired quark matter more stable by
decreasing the energy per baryon. Moreover, we find that the MCFL
matter has an enlarged stability window by comparison with MSQM due
to the quarks' pairing effect. Because of the formation of pairing
quarks and the participation without electrons in MCFL matter, there
is not a drop behavior for the strange quarks even in the strongest
magnetic field. Therefore the MCFL is stable against the transition
of the SQM to non strange quark matter in the neutron stars
\cite{Isayev}.

The anisotropic structure of MCFL matter is also investigated in our
quasiparticle model. The isotropic symmetry in the ordinary CFL
matter is broken by the magnetic field. The anisotropic pressure
splitting is dominated by the magnetic field strength. Importantly,
it is found that the anisotropic structure is almost independent on
the pairing gap, and the threshold value of the magnetic field is
the same for the different phases, i.e., MCFL and MSQM. The result
can be understood for that the contribution of the Maxwell term of
the magnetic field is much higher than that of the pairing effect.

Since the MCFL matter has a larger stability window, one can expect
extensive research of MCFL matter in astrophysics. For example, we
can hope that the CFL and/or MCFL matter can be energetically
favored in neutron stars or pulsars even if ordinary SQM could not
be allowed in the core of neutron star \cite{alford2003}, In
particular, MCFL matter has a larger probability of existing in
stars due to the two mechanisms, namely, cooper pairing effects and
the Landau magnetic quantum levels. The study of magnetic field in
the inner regions of compact stars is helpful to distinguish the
different quark matter phases. The anisotropic equations in
ultrastrong magnetic field are still worth investigating for
non-spherical strangelets and compact stars in future work.

\begin{acknowledgments}
The authors would like to thank support from the National Natural
Science Foundation of China (Grants No.11005071 and No.11135011) and
the Shanxi Provincial Natural Science Foundation (Grant No.
2011011001-1).
\end{acknowledgments}

\end{document}